\documentclass[10pt, conference, compsocconf]{IEEEtran}
\usepackage{epsfig}
\usepackage{mystyle}
\usepackage{law}
\usepackage{url}
\usepackage{comment}
\usepackage{stfloats}

\newcounter{saved}
\def\save{\saveitem{saved}\addtocounter{saved}{-1}}

\def\resume{\stepcounter{saved}\setitem{saved}}
{\begin{rules} \resume}%
{\save \end{rules}}

\begin{document}

\title{Establishing Global Policies over Decentralized Online Social Networks}

\author{\IEEEauthorblockN{Zhe Wang, Naftaly H. Minsky}
\IEEEauthorblockA{Department of Computer Science\\
Rutgers University\\
New Brunswick, NJ, 08903 USA\\
Email: \TT{\{zhewang,minsky\}@cs.rutgers.edu}}
}

\maketitle

\begin{abstract}
Conventional online social networks (OSNs) are implemented in a centralized manner.  Although centralization is a convenient way for implementing OSNs, it has several well known drawbacks. Chief among them are the risks they pose to the security and privacy of the information maintained by the OSN; and the loss of control over the information contributed by individual members.

These concerns prompted several attempts to create decentralized OSNs, or DOSNs. The basic idea underlying these attempts, is that each member of a social network keeps its data under its own control, instead of surrendering it to a central host; providing access to it to other members of the OSN according to its own access-control policy. Unfortunately all existing DOSN projects have a very serious limitation. Namely, they are unable to subject the membership of a DOSN, and the interaction between its members, to any global policy.

We adopt the decentralization idea underlying DOSNs, complementing it with a means for specifying and enforcing a wide range of policies over the membership of a social community, and over the interaction between its disparate distributed members. And we do so in a scalable fashion.
\end{abstract}


\s{Introduction}\label{intro}

An \emph{online social network} (OSN) can be defined broadly
as a \emph{community}  of people that interact with each other via some electronic
media, which generally operates subject to a
\emph{policy} that may regulate the membership of the community, and the manner
in which its members interact with each other.
 The
policy of a  purely social community is
often informal, imprecise,  implicit and only occasionally enforced.
But such policy needs to  be tightened  for  an OSN,
because its membership can be larger than that of a purely social
community, and its members tend to be less familiar with each other.
Therefore, the policy of an OSN needs to be  explicit and well defined,
 and it needs to be
more strictly enforced, largely via computational means, so that it can
establish desired regularities  over the OSN.

Such policies are easily implementable via the conventional
types of OSNs---such as  the currently popular Facebook, Google+, and
Twitter---because of their \emph{centralized architecture}. That is 
each such OSN employs a
virtually \emph{central host}---which may be a centrally managed cluster of
computers---that mediates all interactions between its members, subject 
 to a policy defined by the host.
 This central host also maintains the information supplied by the  members of
 the community in question.
 
Unfortunately although centralization is a very convenient way for implementing
OSNs, it has
 several well known drawbacks, which include:
 (a) lack of scalability;
 (b) the existence of a single point of failure;
 (c) the risks to the security and privacy of information maintained by
 the central host; and (d) the loss of control over the information contributed by
 individual  members.
 The first two of these drawbacks can be mitigated
via very large, complex, and expensive infrastructures---like those used by Facebook and
Twitter. 

But the risk to security and privacy, and the loss of control over private information\footnote{Henceforth,
     we will mostly talk about security, interpreting this term broadly.}  are  harder to
mitigate, because they are mostly the consequence of centralization itself. Indeed, maintaining the state of
the membership  of an OSN, and the history of interaction between members,
under a single administrative domain makes it vulnerable to 
 various malicious attacks. Such
attacks can be mounted by insiders, say the programmer that maintains the
software of the OSN; and by hackers from the outside, for whom the central
repository of information is likely to be very lucrative.

 Security seems not to be of much concern  to the hundreds of millions of current users of Facebook,
Twitter, and similar OSNs. But they are, or should be, of serious concern to
other types of OSNs, whose members exchange more sensitive information---such
as private medical and financial information; and information about the
business of an enterprise, exchanged between its employees.
We will consider  examples of such OSNs in the following section.

Such concerns about centralized OSNs prompted several attempts to create
decentralized OSNs, or DOSNs; such as LotusNet \cite{aie10-1}, Safebook \cite{cut09},
PeerSoN\cite{bod11}, and others.
The basic idea underlying all these attempts to the decentralization, is that
each member of the community in question
should keep its data under its own  control, instead of
surrendering it to a central host, providing  access to it to other members of
the DOSN according to its own access-control policy.

Unfortunately all existing DOSN projects have a very serious limitation. Namely, 
they are unable to subject  the membership of a DOSN, and the interaction
between its members,
to any global policy. This is a very serious limitation of the DOSN
architecture, because, as pointed out above,  an enforced global policy is 
generally essential for an OSNs, as it helps make it into a social community.

\p{The Contribution of this Paper} We will adopt in this paper the 
 decentralization idea underlying DOSNs, complementing it with a
 means for specifying and enforcing  a wide range of policies over the membership of a social
community, and over the interaction between its disparate distributed members.
And we shall do so in a scalable manner.

 The rest of this paper is organized as follows.
 \secRef{examples} introduces examples of OSNs for which security 
is critical, and would thus benefit from decentralization.
 \secRef{regularity}  introduces examples of policies that are often
essential for an OSN---particularly for the types of OSNs for which security
tends to be critical---but which cannot be established under DOSN.
 \secRef{lgi} provides a very brief outline of the LGI middleware, which serves
 as the basis for this work.
\secRef{model} introduces our model of decentralized OSN---we call it OSC, for
``online social community,'' where the term ``community'' is meant to suggest
two things: first, a decentralized nature, like that of purely social
communities; and second, the existence of a shared policy, which characterize
most  social communities, and which under OSC is
enforced.
\secRef{case} is an implemented case study that demonstrates how this abstract model
 can be used for a concrete application. The related works are discussed in \secRef{related}.
And we conclude in \secRef{conclusion}.

\s{Examples of OSNs, for which Security is Important}\label{examples}
We distinguish here between two types of OSNs: (1) \emph{autonomous
OSNs}, which are not bound by any outside authority; and (2) \emph{bound
OSNs}, which operate in the context of some organization, which has
jurisdiction  over it. We focus in this section on the security needs  of these
two types of OSNs, and on the risks to security that centralization poses to them.
We will discuss both types of OSNs, but we will focus, here 
and in the rest of the paper,  on the latter one.

\ss{Autonomous OSNs}\label{consult}
Consider a set of physicians who form an OSN that enables them to 
 consult with each other about various medical issues they confront.
This \emph{MD-Consultation} OSN is to admit only qualified MDs as members, and
may grow to be quite large if physicians all over the world join it.
The information exchanged between members of such OSN is clearly very
sensitive.

So,  having the consultation process mediated by a central host, and having the
information exchanged between the physicians maintained centrally by this host
 can seriously compromise the security
of both the doctors and their patients.
The risk here is particularly serious 
because the host of such an OSN is likely to become a target for  attackers, since
the information maintained by it can be exploited
for illicit financial gains. 

 There are many potential  autonomous  OSNs that exchange similarly
sensitive information; such as an OSN that enables  people who suffer from a
certain malady, to share their experience with each other, and with their
doctors;  an OSN formed by a certain type of workers for exchanging their
views about their employers;  an OSN used by students to exchange information
about their teachers, and many others.

\ss{Bound OSNs} \label{workplace} 
There is a growing realization\cite{zha09}
that  OSNs that operate within an organization---such as
 manufacturing,  commercial enterprises, medical centers, or
even the military---can be beneficial for it. This seems to be particularly the
case for OSNs that provide for micro-blogging, as  is evident from the
 recent purchase of the
Yammer---a prominent micro-blogging OSN operating within
organizations\cite{yammer}---by Microsoft,
for \$1.2 Billion.  We will have more to say about Yammer itself, but first we
outline  some of  functional features one can expect from this kind of OSN.

 Consider 
 a large and geographically distributed enterprise $E$ that  provides a
 centralized micro-blogging OSN  for its employees.
Suppose that such an OSN, which we call $B_E$,  distinguishes between groups of employees, enabling the
members of each groups  to communicate with each other.
Such groups may be the following: 
 (a) all the  employees of $E$;
(b) the non-managerial staff of $E$;
(c) the managerial staff of $E$;
and (d) members of various task forces operating in $E$.
Note that these groups may overlap partially, as a single employee may belong to
several groups.

Let the members of $B_E$ hold a profile, which is a set of attributes of each
member, which are visible to the whole OSN and can be indexed and searched.
They communicate mostly via micro-blogs by means of some form of a publish/subscribe
(P/S) mechanism.  When using P/S, members can publish
\emph{posts} and build subscription relationships with each other, in some
analogy to the \emph{following} relationship in Twitter.  We assume that each post contains
two parts: \emph{type} and \emph{body}. The type is denoted by using
\emph{\#type\#} at the beginning of a post.  Besides publish/subscribe based
communication,  members can send
direct messages to each other, which we assume here to  be preceded by a type
field, like a post.

\p{The Need for  Security} 
We will distinguish here between two types of needed security.
First, the information exchanged between the employees of enterprise $E$ can
carry sensitive information about the business of this enterprise. It is
therefore important for this information not to be
exposed to the outside, at least not on a large scale.

Second,  one may need to prevent information exchanged between the members of
a certain  group of $B_E$ from being accessible to anybody else, or  to certain other
groups. For example, suppose that enterprise $E$ has several task forces that
consult to other companies, some of which may compete with each other.
And suppose that the members of each such task force form a group in $B_E$.
It is obviously  paramount for these subgroups not to have access to each other's
information. 

\p{The Risks to Security due to Centralization}
There are two types of centralization to be considered, which we call strong
and weak. Strong centralization is like the one
 practiced by Yammer, the Microsoft OSN that we mentioned
above.
Yammer provides  services to a host of different enterprises---they currently
claim to serve  about 200,000 of them. Yammer supports policies that provide
necessary separation between  the various  enterprises it serves. But the
information belonging to all
these enterprises is maintained centrally by the Yammer system. Such centralization of commercial and industrial information of many
companies, is likely to attract attacks from the inside of Yammer, and from the
outside. 

A better approach  would be to  use an intramural  Yammer-like OSN. This,
weaker form of centralization,  would be
much safer than using Yammer. But if this system relies on a centralized
database, it would  still be vulnerable to breaches of security. Indeed, if all the information generated by
the $B_E$ is available to its software, then the  rogue programmers of this OSN
will have a fairly free access to all of it, disregarding the required
boundaries between different groups.

\s{A Sample of Policies that an OSN may Need to Enforce}\label{regularity} We
illustrate here the type of \emph{communal policies} that an OSN may need to
establish. By ``communal'' we mean either global policy that is to govern all
members of an OSN, or a policy that governs some subgroup of its members.
All the policies discussed here can be easily established by a centralized OSN,
but none of them can be established under the  DOSN architecture.

We will distinguish here between three types of communal policies, and will
motivate some of them in the context of the $B_E$ example OSN, introduced in
the previous section. We will show in \secRef{case} how such policies can be
realized in decentralized OSNs. 

\ss{Membership Control} Control over membership is crucial to many social
communities, whether it is autonomous or bound.
Such control may have several complementary aspects. We will consider three of
these below.

First, one may require that to be a member of a given OSN one needs
 to authenticate itself
via a specified kind of certificate. One may think that this policy can be
established under the DOSN architecture
by having every member of the DOSN in question  require every interlocutor of
his to
authenticate itself in a specified manner. But DOSN has no way for ensuring
that all its members behave in this way. (The inability of DOSN to enforce
communal policies is even more obvious in the rest of our examples below.)

Second, one may require that to be a member of an OSN one needs
to  garner the  support of several (say three) current
members of it.

Third,  it is often important to establish some procedure for removing  members
from a given OSN. This can be done in many ways. For example, consider a
OSN that has a member that plays the role of 
a \emph{manager}. Now, let the manager be given the power to  remove any
current member $x$ of the OSN, simply by sending it a 
 message  \emph{remove}. Then  $x$ should lose its ability to interact with
 other members of the OSN.

\ss{Constraints on the Behavior of Members of an OSN}
Sometimes one needs to impose constraints on what members can do. Such
constraints may depend on the profile of individual members, and on the history
of their interaction with others. We have just  seen  an example of such
constraints: only a member that plays the role of manager can send the
\emph{remove} message to others. And any member that gets such a message must
cease all communication with others.

More generally---but stated in the context of the $B_E$ OSN---the
 type of messages that members are allowed to send, or the type of posts
 that they are allowed to issue, may depend on their roles in this OSN,
 which may be represented by their profile.
As another example, a member should be able to force an interlocutor to reduce
the frequency of messages it is sending to it, or to cease sending messages to
it altogether.

\ss{Global Access Control (AC) Policies} One of the intended consequences
of decentralization under DOSN is that it enables each member to apply its own
AC policy to its own data---e.g., to the set of posts it produced, which are
maintained in its own database.  The problem with this aspect of DOSN is that,
unlike the case of Facebook or Twitter, a member of an OSN may not have the
complete authority over the data it maintains.  A case in point is a bound OSN,
such as  the $B_E$
OSN introduced in \secRef{workplace}. The posts being produced by the
various members of this OSN really belong to the enterprise $E$, which
thus has the ultimate authority about how they should be distributed. The
enterprise may relegate to individual members the right to apply their own AC
policies, \emph{provided} that these policies conform to the global policy of
an enterprise. For  example, the global policy of
the $B_E$ may be that a group of members assigned to deal with the business of
a given client-company
 can communicate only with each other, as long as they operate as members of
 that group---recall that under $B_E$, a single person may belong to several
 groups.

\s{The Law-Governed Interaction (LGI) Middleware---an Overview}\label{lgi}
 LGI is a middleware that can govern the interaction (via message
exchange) between  distributed \emph{actors}, by 
 enforcing an explicitly specified law---and possibly multiple laws---about 
such interaction. We provide here a brief, and rather abstract, overview of LGI; focusing on what is
 the most relevant to this paper.  A more detailed
presentation of LGI, and a tutorial of it, can be found in its manual
\cite{min05-8}---which describes the release of an experimental implementation
of the main parts of LGI.  For additional
information and examples the reader is referred to a host of published
papers, some of which will be cited
explicitly in due course.

The rest of this section is organized as follows.
We start, with the local nature of the interaction laws
under LGI---a key characteristics of this middleware that enables many of the
novel features of it.
We then discuss the following aspects of LGI:  the structure of its laws;
and the law enforcement mechanism.

\ss{The Local Nature of Interaction Laws}\label{local}
 Although the purpose of interaction laws is to govern the exchange of messages between
 different distributed actors,
 they do not do so directly under LGI. Rather, an LGI law \EL\ governs  the
 interaction   of any actor operating under it, essentially by
 controlling  its ability to send messages to others, and to
 receive messages from them\footnote{In fact, a law can also  cause  messages
 to be changed and rerouted,  and it can  change the state of an agent.}. 
 A law  \EL\ is local to each actor $x$ operating
 under it, in  that  its rulings are based  solely on the local state of $x$
 and on the  event that occurs at it,  and are completely  independent
 of the coincidental state  and events occurring anywhere else in the system. Such a law can be
 enforced locally, and thus very scalably, in a manner described in
 \secRef{enforcement}.
Moreover, the locality of LGI laws has several other beneficial consequences,
some of which will be pointed out in due course.

It should also be pointed out that although locality constitutes a strict
constraint on the structure of  laws, it does not reduce their expressive
power.
This has been proved in
\cite{min05-8}. 
In particular, despite its \emph{structural locality}, an LGI law can have
\emph{global effect} over what is called an  \CAL{L}-\emph{community}, defined
as the set of  actors operating under a common law \EL.

\ss{LGI Laws---a Definition}\label{law-def}
An  \emph{interaction law} (or simply a \emph{law}) \EL\ is defined over three
elements---described with respect to a given  actor $x$  that operates under
this law:  (1) a set $E$ of \emph{interactive events} 
that may occur at any actor, including the arrival of a message at $x$, and
the sending of a message by it;
(2)  the \emph{state} (also called the  \emph{control-state}) $S_x$  associated
 with each actor $x$, which  is distinct from the internal state of $x$,
that is   invisible to the law;
 and (3) a set $O$ of
 \emph{interactive operations}
 that can
 be mandated by a law, to be carried out at $x$ upon the occurrence
 of interactive events at it;
this set includes  operations that forward messages to others,
along with some other types of operations that have an
effect on the flow of message into  $x$ and from it.

Now, the role of a law under LGI is to decide what should be done in response to
the occurrence of any interactive event at an actor operating under this law.
This decision, with respect to actor $x$,
is defined by the following mapping:
\begin{equation}
 \mathcal{L} : E \times S_x \rightarrow  S_x \times (O)^*
\label{eq-law-2}
\end{equation}
In other words,  for any  a given  $(event,state)$ pair, the law mandates a new
state (which may imply no state change), 
as well as a (possibly empty) sequence of  interactive operations.
 Note, in particular, that
the ruling of the law at a given moment of time depends on the state of $x$ at
that moment; and that the 
evolution of the state itself is determined by the law, and by the history of
interactive-events at $x$. LGI laws are, therefore, stateful, and sensitive
to the history of interaction.

Note that the law is a complete function, so  that any mapping defined by Formula~\ref{eq-law-2} is considered a valid
law---which  means that a law of this form is \emph{inherently self consistent}. This
does not mean, of course, that a law cannot be wrong. It can be wrong in
 the sense that 
 it does not work as
intended by its designer; but this is not a matter of inconsistency.

Finally, it is worth pointing out that while  Formula~\ref{eq-law-2} is a 
definition of the semantics of laws\footnote{Modulo the fact  that  the sets $E$ of events and $O$ of
operations have not been fully spelled out here.}, it does not
 specify a language for
writing laws.
 In fact, the current implementation of LGI supports two different
\emph{law-languages}, one based on the logic-programming language Prolog, and
the other based on Java. But the choice of language has no effect on the
semantics of LGI, as long as the chosen language is sufficiently powerful to
specify all possible mappings defined by Formula~\ref{eq-law-2}.

\ss{The Decentralized Law Enforcement Mechanism}\label{enforcement} 
Consider an actor $x$ that chooses to operate under a law
\EL. It can do so by \emph{adopting} a generic controller as its mediator, loading law
\EL\ into it. Once thus adopted, this controller is
denoted by   $ T^{\mathcal{L}}_{x}$---meaning that it operates under law \EL,
serving actor $x$---and  the   pair $\langle x, T^{\mathcal{L}}_{x}\rangle$, is
called \emph{agent} $x$ and is referred to as  an
\EL-\emph{agent}---and sometimes  simply an ``agent''. This adoption, which   signifies the \emph{birth} of agent
$x$,   is one of the interactive events of LGI, so that the law in question has
the possibility of refusing to be adopted by this actor, and can
mandate  some initialization  for it, if it does not refuse.

Note the fundamental difference between a bare actor and its agent: while
the interactive behavior of an actor  is unpredictable---unless its code is
 known---the interactive behavior of an
\EL-agent is known to  conform to  law \EL.

\begin{figure}
\leavevmode
\epsfysize=0.8 in
\centerline{\epsffile{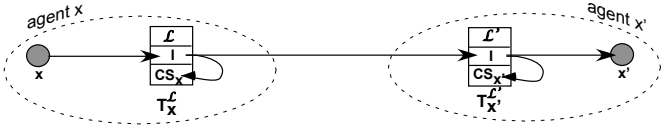}}
\caption{Interaction between a pair of  agents, mediated by a pair of
controllers under possibly different laws.
}
\label{fig-agent}
\end{figure}

 \figRef{fig-agent} depicts the manner in which a pair of agents, operating
 under possibly different laws, exchange a message. (An agent is depicted
 here by a dashed oval that includes an actor and its  controller.)
Note the  \emph{dual nature} of control exhibited here:
the transfer of a message is first mediated by the sender's controller, subject
to the sender's law, and then by the controller of the receiver, subject to its
law. This dual control, which is a direct consequence of the local nature of
LGI laws,
 has some important
consequences. In particular, it facilitates flexible interoperation and it enables more sophisticated control than possible
under many AC mechanisms that provide control only on the receiver side.

The overhead incurred by this kind of control turns out to be relatively small.
In circa 2000 it was measured to be around 50 microseconds for fairly common
laws, which is  negligible for
communication over WAN. This is one of the results of  a comprehensive study of
this overhead in \cite{min99-5}.

Finally, we note that a generic controller needs to be trusted to enforce
correctly any law it is adopted with.  There are several ways for providing such
trusted controllers as the TCB (Trusted Computing Base) of the system in question.
In the case of a bound OSN, like our $B_E$ example, we
expect this to be done by the enterprise $E$, in the context of which $B_E$
operates.
This company could construct
what is called a  \emph{controller service} (CoS)
that maintains a set of well tested \emph{controller pools}, each of which can
host a number---it is usually in the hundreds---of individual
controllers that can be used by arbitrary
actors, upon request.
For other types of OSNs one expects the CoS to be maintained by some commercial
company that provides its services for a fee.

Note, therefore,  that a controller    $ T^{\mathcal{L}}_{x}$ and the actor $x$
that adopted it would run on different hosts. This would help prevent $x$ 
corrupting its own controller.
Even if a controller is hacked, since it does not keep the messages it passes, there is no way to get the information of the whole history.
And since it would be much harder to compromise many controllers than one, the global view of the whole system will not be obtained.

\s{A Model of Decentralized OSN}\label{model}

We introduce here a model of decentralized OSNs that differs from the current approach to
the decentralization employed under the current DOSN architecture,
 in that it enables the enforcement of 
communal policies over it. 
 We call a specific OSN under this model  an  \emph{online
social community} or an OSC (or sometime simply a  \emph{community}), and
we refer to this model itself as the OSC-model.

Now,  a community $C$  under the OSC  model
 is broadly defined as a 4-tuple
 $\langle M,$ \EL, $T, S\rangle,$ 
where $M$ is the set of members of $C$;
\EL\  is the policy that governs this community, which we call a
\emph{law} (an LGI-law, to be exact);
 $T$ is a set of generic LGI controllers that serve as the middleware
that enforces law \EL; and
$S$ is  a set of components
that support the  operations of $C$, and is specific to it---this set is
called the \emph{support} of $C$, and it may be empty.

We now elaborate on this schematic definition of the OSC model by discussing the
following aspects of it: (1) the anatomy of a community under OSC; (2)
the launching of  an  OSC-community;   (3) the
operations of a  community; and (4) possible extension of this model.
Note that an example of  an OSC-community is described in \secRef{case}. 
\begin{figure*}
\leavevmode
\epsfysize=3.8 in
\epsfxsize=4.2 in
\centerline{\epsffile{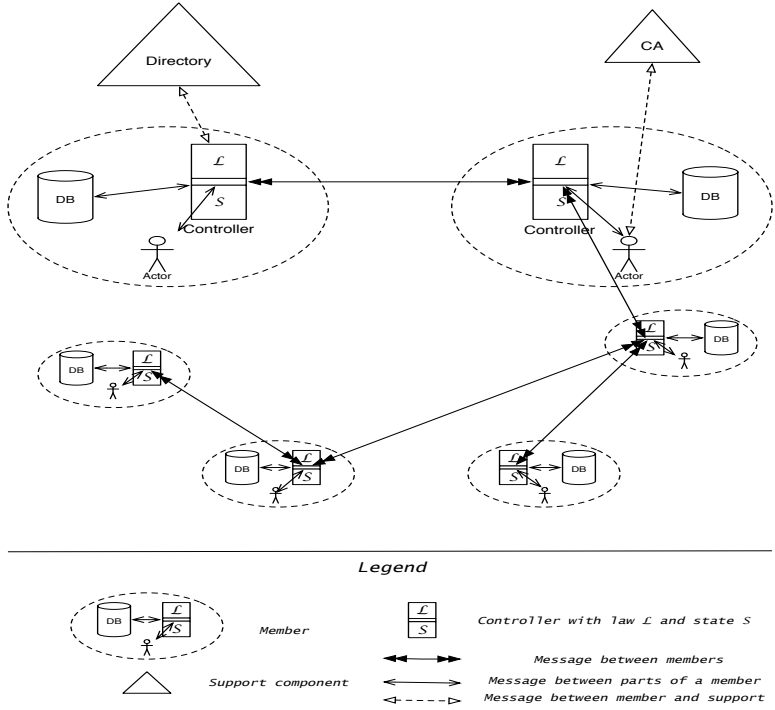}}
\caption{The Anatomy of an OSC Community
}
\label{fig-arch}
\end{figure*}

\ss{The Anatomy of  a Community Under OSC}\label{anatomy}
We describe here the anatomy of a community $C$ under this model by elaborating on its various
components, and on the relations between them. This anatomy is depicted 
schematically in 
 \figRef{fig-arch}.

\p{The Set $M$ of Members}
An individual member $m$ of   a community $C$ is a triple 
 $\langle user, mediator, database\rangle,$ 
where \emph{user} is usually a human, operating via some kind of computational
platform, like a smart phone; \emph{mediator} is an LGI-controller that mediates all interactions
between this member and the rest of the community---as well as between the
other two 
components of the member  in question---subject to law \EL $C$ (which we
denote by \law{C}); and \emph{database}, which is an optional part of the
member,
 is the private database of $m$ that maintains
 information associated with this member, such as the set of Twitter-like micro-blog posted by $m$,
 or its Facebook-like page.
This database is meant  to be controlled by the user, and maintained either on its
own host, or on some cloud. (Note, however, that  a community that operates
within an organization may require the databases of members to be maintained
somewhere in the Intranet of this  organization.)

\p{The Law \law{C} of community $C$}
It is the law which
endows an OSC-community with its overall structure and behavior.
And the fact that the law can, in principle, be any well  formed LGI 
law (cf. \secRef{lgi}) endows this model with great deal generality regarding
the policy that can be enforced over a community.

\p{The set $T$ of LGI Controllers}
Every user can create its own controller, using the software provided by the
released  LGI middleware. But if malicious corruption of  controllers by their
users is of concern, then
it is better  for the members of a community to adopt controllers created and maintained  by a trusted  \emph{controller service} (CoS),
so that they  can
authenticate each other as bona fide LGI controllers.
For such a CoS to be trusted to provide genuine controllers, this service needs to
be managed by a trusted organization. In particular, the CoS may be managed by
the organization in the context of which the community is to operate---as in the case
discussed in \secRef{case}. 
Alternatively,  the CoS may be created and
managed for general use, by a reputed organization 
which has no interest in the applications that use its controllers.
Such applications can be any kind OSC-community.
For  more about the
security and trustworthiness of controllers see  \secRef{lgi} and \cite{min05-8}.

\p{The Support $S$} 
An OSC-community    may require services of various components that
are not themselves members of this community.
Here are some examples of such components:
 (a) a certification authority (CA) used for the authentication
 the various members of the community; (b)  a \emph{naming service} that provides unique
 names of community members; 
(c) an index service for searching; and (d) a networking service for
    maintaining various networking structures of the community---more about
    which in \secRef{discussion}.
It is worth pointing out that this  set of support components
may  be empty for some communities.

\ss{The  Launching of an  OSC-Community}\label{deploy}
A specific  OSC-community, $C$ 
is launched by constructing its \emph{foundation}---described below---and then having individual
members join it.
The construction of the foundation of a community $C$ consists of the
following steps: (a) defining law \law{C}   under which  this community is to operate;
 (b) implementing the required
support components; and (c) selecting,
or constructing, a
\emph{controller-service} (CoS) for the use of this community---or providing
means for prospective members to construct their own, TPM-based, controllers.

Once the foundation of $C$ is constructed, anybody can attempt to join
it as a member, via the following three steps: (a) deploying its private database---if
one is required by law \law{C}---with
an API required by this law;
(b) adopting an LGI-controller, and loading  law \law{C} into it; and (c) providing
this controller with a pointer to its database, if any.
It should be pointed out that  such an attempt to join a given community $C$ may
fail, if the conditions for joining imposed by law \law{C} are not satisfied.

\ss{The Operation of a Community}\label{operation}
Consider a member $x$ of a community $C$ sending a message $m$ to another member $y$. The message first
arrives at the controller of $x$, that operates under law \law{C}.
These controllers would then carry out the ruling of law \law{C}, which can
mandate the execution of   any number of the following kind of actions: 
(a) change its own state in some way; 
(b) communicate with the database of $x$;
(c) send the message $m$, or some other message, to the controller of the
    original target of $x$; and
(d) send some other messages to the controllers of some other members, or to
    some of the support components of the community.
Among other things, this means that
 members of a community interact with each other via
their controllers, and the controllers communicate with each other. 

 It is
worth pointing out here that LGI provides an important \emph{trust modality} which
is critical to this model. This trust modality is called
\emph{law-based trust}, or simply \emph{L-trust}, and  can be introduced, broadly,
as follows:
any pair of  interacting LGI-controllers can identify, cryptographically, each other
  as  genuine
controllers, and can identify the law, under which their interlocutors operates.
One consequence of this  is that the law \law{C}   of the given community $C$
 can be written so that members of $C$ can interact only with other members of $C$.
Now, \emph{L-trust} can be defined as follows:
 \emph{members of a community $C$ can trust
each other's interactive behavior to comply with their common law \law{C}}. For a complete
definition of this trust modality and some of its consequences see
\cite{min12-2}. 

Another important  observation about the behavior of a community under this model needs to
be made:
 the ruling of a law  for a given event that occurs at a
controller depends on the state of this controller, which may be different for
different members.
This difference can come from some certificates submitted by the user to its
controller, which may authenticate the role of the user in the organization in
question. And the state may change dynamically in response to some interactive
activity of the community. For example, the manager of the community under our $B_E$
community, may be allowed by the law of $B_E$ community to transfer its managerial baton to
some other member, which would then be able to send \emph{revoke} messages.
In other words, \emph{the members of a  community $C$ may not be equal under its law
\law{C}}.

\ss{Discussion: on Networking and on Scalability}\label{discussion}
We have already pointed out that some capabilities are easier to be provided via
centralized OSN than via decentralized one. We have focused on
the imposition of communal policies over an OSC in this paper. Another capability that is
problematic under decentralized OSNs is the ability to  analyze the networking
relationship implicit in the community. Consider for example the \emph{friend}
relationship of Facebook, and let us examine its realization in an OSC.

It is easy to have each member of an OSC list his friends---we have done
with a similar relation in \secRef{case}---but it is very hard and expensive to
analyze the entire friendship-graph, when this relation is recorded in such a
distributed manner. Of course, such global analysis, which is central to
Facebook, is not required for all kinds of OSN. But it is often required, and
must be provided for.

A reasonable way for enabling global analysis of a network implicit in an OSC,
is to maintain it explicitly in a central manner. That is, we maintain the
friendship relation (or any other kind of relation between members) in a
central place, as part of what we have called the \emph{support} of an OSC, and
then provide these components with various analysis tools.
This is a reasonable solution under two conditions: (a) the relation in
question is not, itself, highly sensitive from the privacy and security
viewpoint; and (b) the central \emph{network component} is not used too
frequently, so it would not reduce substantially the scalability of the OSC in
question.

More generally, an OSC may have several centralized support components, such as
indices of various kinds. If these components are not used very frequently they
would not seriously undermine  the scalability of an OSN, due to the
decentralization of its data and of its policy enforcement mechanism.

\ss{Towards an Extension of this Model}\label{extension}
We have seen in \secRef{examples} that a social community may have several
groups, or sub-communities.
All such groups may operate under a single law, as demonstrated in
\secRef{case}. But such a single global law may be hard to design, hard to
reason about, and inflexible with respect to changes of the OSC.

These problems can be alleviated  via the concept of \emph{conformance
hierarchy} of LGI laws \cite{min03-6}. Using this concept, an OSC can be built to be governed
by a tree of laws. The root of this tree, say \law{R}, would govern the entire
community, while each sub-group $g$ of that community would be governed by  a
law \law{g} defined as  subordinate to \law{R}, and is thus constraint to
conform to it. This way of governing an OSC is very modular and flexible; and
it will be described in a forthcoming paper.

\s{A Case Study}\label{case}

In this section, we describe the implementation of the $B_E$ community,
introduced in \secRef{examples}. It has been implemented in the scale of more than two hundred users as a proof of concept.
This community operates in the context of  a large and geographically distributed
enterprise E,  providing
a micro-blogging OSN for its employees, as a complement to its existing office
systems. $B_E$ enables the members of various groups of employees to communicate
with each other. The groups of this community are (1) all employees; (2)
management staff; (3) non-management staff; (4) members of task force
\emph{t1}, which is providing consultation service for an enterprise \emph{E1};
(5) members of task force \emph{t2}, which is providing consultation service
for another enterprise \emph{E2}. These groups could partially overlap, in the
sense that a single employee may belong to several groups.

As described in \secRef{examples}, there are two modes of communication in this community: publish/subscribe and
direct message. And each post or message contains two parts: type and body. 
For both communication modes, there are certain global
policies can be imposed to the community to control members' behaviors. We will
discuss them in the following section.
 
Each member of the community holds a profile in its controller, as well as
several internal states, which are used for some functionalities and not
visible for regular members. A profile is a group of attributes of the user,
which are visible to the whole community and can be searched and indexed. There
are mainly two types of attributes in the profile. One type of the attributes
is the relatively stable attributes, like real name, login ID, position, group
identity, age, etc. These items usually require users to provide certificates in
order to get them in the profile by the rule of authentication. Since these
attributes are stable, an index for searching is able to be built on
them. Another type of attributes is the dynamic ones, such as interest, skill
set, last ten posts, etc. Although these items don't require certificate, not
all of them can be changed by the member arbitrarily. For example, subscriber
list is handled by the subscription mechanism, reputation is maintained by
controller according to the rates gotten from other members and an attendance
attribute could be decided by the sign-in/sign-out time. We call these
user-unchangeable dynamic attributes and the certified attributes together as
controlled attributes, and the rest attributes as discretion attributes. The
internal states are the states maintained by the controller for certain functions
of the community. For instance, the frequency of publishing is used for
preventing a member overwhelming the community by violently publishing posts.\\

 \noindent \textbf{The Law of the $B_E$ Community:} \\
 
 The law \CAL{B} of the $B_E$ Community is used for regulating every aspects of the operations and behaviors of the community. We split it into several parts according to their functionalities. Due to lack of space, we only discuss the detailed law of some functionalities of the communities. In \secRef{member}, we discuss how a user becomes a member of the $B_E$ community and its groups, how it configures its profile, and how a member is removed. \secRef{communication} shows the communication mechanism and the imposition over it. We discuss other functionalities which are needed to be a complete OSN in \secRef{others}.

 \ss{Member Profile and Membership Control}\label{member}

To join the community, a member needs to adopt a controller under law \CAL{B}. Rule~\ref{rule-member-employee} allows a user to join the community by presenting a certificate from a CA run by the enterprise in question to prove that it is an employee. Once certificate is verified by the controller, the set of attributes in its profile will be inserted into the user's control state. An example of an attribute is \emph{role(manager)}. There are two types of attributes in the profile: certified and uncertified. The certified attributes are the relatively stable ones, like real name, login ID, position, age, etc. These items can also be obtained by providing certificates after the adoption. Rule~\ref{rule-member-group} allows the user to join the group $t_i$ by providing a group certificate. It will add an attribute $t_i$ to its profile, as well as an access control filter which we will discuss in the next section. A database access API is provided for members. It supports CRUD (create, read, update and delete) functions for users to access their database. When adopting a controller, a member is required to provide the address of the database to associate with its id. In our example, the member can only provide the address with the enterprise domain so that the enterprise has the physical access to it and can employ firewall to protect it.

\begin{ruleset}{Law \CAL{B}: Member's Profile and Membership Control \label{law-member}}
\begin{footnotesize}

  \Rule UPON adopted(X,cert(issuer(ca),subj(X),attr(A))):- \\
  do(+A).\label{rule-member-employee}
  
  \Rule UPON certified(X,cert(issuer(ca),subj(X),attr($t_i$))):- \\
  do(+$t_i$);\\
  do(+filter(group($t_i$))).\label{rule-member-group}
  
  \Rule UPON sent(X,addProfile(Attribute(Value)),X):- \\
  if ( $\neg$ (Attribute in controlledAttributes) )\\
  then do(+Attribute(Value)).\label{rule-member-addattribute}

  \Rule UPON sent(X,updateProfile(Attribute(Value)),X):- \\
  if ( $\neg$ (Attribute in controlledAttributes) )\\
  then do(-Attribute); do(+Attribute(Value)).\label{rule-member-updateattribute}

  \Rule UPON sent(X,addFilter(Attribute(Value)),X):- \\
  do(+filter(Attribute(value))).\label{rule-member-filter}

  \Rule UPON sent(X,\#revoke\#,Y):- \\
    if(role(manager)@CS) then do(Forward);\\
    else do(Deliver(X,notAllow,X)).\label{rule-member-sendrevoke}

  \Rule UPON arrived(X,\#revoke\#,Y):- \\
  update(certificateBlacklist);\\
  inform(certificateBlacklist);\\
  do(Quit).\label{rule-member-receiverevoke}
    
 \Rule UPON sent(X,\#db\#M,X):-\\
 if(M in CRUD)\\
 then do(Release(X,M,DB)).\label{rule-member-db}
 
 \Rule UPON submitted(DB,Q,X):-\\
 do(Deliver).\label{rule-member-dbr}
    
  }\save

\end{footnotesize}
\end{ruleset}

Another type of attributes is the dynamic ones, such as interest, skill set, last ten posts, etc. Although these items don't require certificate, not all of them can be changed by the member arbitrarily. We call the user-unchangeable dynamic attributes and the certified attributes together as controlled attributes, and the rest attributes as discretion attributes. For the discretion attributes, user can directly add some of them into its profile via Rule~\ref{rule-member-addattribute} and update them via Rule~\ref{rule-member-updateattribute}. Rule~\ref{rule-member-filter} shows how a user sets up its subscription filter. Sometimes the user may not want to be subscribed by everyone. The existing way to do that in other social networks is to put the subscriber into blacklist, or we say to block specific user. This can only happen after somebody initiated or requested the subscription and needs to be done manually. However, our mechanism can prevent
subscription by specifying a certain kind of attributes. User can use Rule~\ref{rule-member-filter} to add the filter content into control state. Its controller will only allow the members who have the required attributes to subscribe to it. The following operations will be described in \secRef{communication}.

Finally, rules~\ref{rule-member-sendrevoke} and \ref{rule-member-receiverevoke} regulate the removal of members from the community. Rule~\ref{rule-member-sendrevoke} shows that only the manager role can remove a member from the community. Non-managers are not allowed to use the type \emph{revoke} when sending messages. When the \emph{revoke} message arrives at the member's controller, according to Rule~\ref{rule-member-receiverevoke}, the controller will directly terminate the connection to the actor and then put its certificate to the blacklist and broadcast to all controllers. Next time when another actor tries to use this certificate to adopt a controller, the controller will not verify it. The member has no way to control or avoid that. This rule guarantees that its participation in this community is seized immediately after the manager removed it and cannot get back again using the same certificate. This is just an example of how to handle the membership removal. Other methods, such as suspension, can also be supported. Rule~\ref{rule-member-db} shows the API provided for the database access. When the member sends a CRUD message, the controller will forward the query to its database. If the query is a Read request, the controller will deliver the query result when the database replies, as in Rule~\ref{rule-member-dbr}.

\ss{Communication}\label{communication}

There are two modes of communication in this community: publish/subscribe and direct messaging. By P/S, members can publish \emph{posts} and build subscription relationships with each other, in some analogy to the relationship \emph{following} in Twitter. For a member to subscribe to the posts from another member, the subscriber \emph{s} sends a subscription request to the publisher \emph{p}. When \emph{p} receives the request, it will add \emph{s} to its subscriber list unless such subscription is prohibited by the law, or if \emph{p} itself blocks the subscription. When a post is published by a member, it will be automatically pushed to all its subscribers. Moreover, members can also send direct messages to each other, which can also be controlled.

We will show later in this section, how the communication is enabled and controlled. The control over communication has two complementary parts: global control and local control. The global control is imposed on every member of the community, but can be sensitive to the state of members, while the local control is discretionary to each member. We discuss both of controls below, and the according law later.

\p{Global Control}\label{global}

The global control over publish/subscribe is imposed on both publishing and subscription. The control over publishing is on what types of posts a member can publish. For example, only the managerial staff can publish posts with type \emph{management}. Upon publishing, a management post is allowed to be published only when the member has the attribute \emph{role(managerial)} in its profile.

The control on the subscription regulates who can subscribe to whom, and to which types of posts. Essentially, it is defined by a condition $C$ on the profiles of the publisher and subscriber. An example of such global policies is that only the members from a same group can talk to each other. The problem is that there is no single place where these profiles can be evaluated because of the decentralization. To solve this problem, our law forces every subscription request to include the profile of the subscriber. And then it has the condition $C$ to be evaluated and acted upon at the publisher side. This can be achieved by checking the profiles of the publisher and subscriber and rejecting the subscription request if the two members are from different groups.

The control over sending direct message is similar to the one over publishing. Certain types of direct messages are allowed to be sent only when the members have the required attributes in their profiles. For instance, only the manager role can send the \emph{revoke} message.

The control on receiving the direct messaging is different from the one on subscription. Whenever a member sends a direct message, its controller will append its profile to the message. Upon the arrival of the message at the receiver side, the controller of the receiver will not deliver the message if its profile does not satisfy the condition of receiving it.

\p{Local Control}\label{local}
Sometimes, a member does not want to be subscribed by certain members, it can block the subscription requests from them. To achieve this, we introduce a profile attribute called \emph{filter}. If a member adds a filter \emph{filter(X)} in its profile, its cannot be subscribed by the member who has attribute X in profile. As we described above, whenever a subscriber \emph{s} sends the subscription request to a publisher \emph{p}, it will be forced to attach its profile along with the request. When the request arrives at the publisher's controller, \emph{s} will not be added to \emph{p}'s subscriber list if its profile has the banned attributes in the filter. This rule of filter is just an example. More complex uses of the filter, such as OR or XOR, could also be achieved.

\p{The Law}
The rules of law \CAL{B} that implements these provisions are defined in Figure \ref{law-communication} and described below.

\begin{ruleset}{Law \CAL{B}: Communication\label{law-communication}}
\begin{footnotesize}

\resume
  
  \Rule UPON sent(X,publish(P),X):- group($t_i$)@CS\\
  if (typeof(P) == \#management\# and $\neg$ role(manager)@CS)\\
  then return;\\
  updateProfile(lastTenPosts(P)); \\
  updateDB(P);\\
  if(subList[group($t_i$)] = []) then return; \\
  else forEach(subscriber in subList[group($t_i$)]) \\
  do(Forward(X,P,subscriber)).\label{rule-communication-post}

  \Rule UPON arrived(X,P,Y):- \\
  do(Deliver); \\
  do(inform(X,P,Y)).\label{rule-communication-receive}
  
  \Rule UPON sent(X,requestSubscribe(profile),Y):-\\
  do(Forward).\label{rule-communication-requestsubscribe}

  \Rule UPON arrived(X,requestSubscribe(profile),Y):- group($t_i$)@CS\\
  if(filter(Attribute(Value))@CS and Attribute(Value)@profile)\\
  then do(Forward(Y,subscribeNotAllowed,X));\\
  else do(updateSublist[group($t_i$)]); do(Forward(Y,subscribeAllowed,X)).\label{rule-communication-subscribereply}
  
    \Rule UPON sent(X,M,Y):-  \\
  do(Forward(X,M(profile),Y)).\label{rule-communication-dm}
  
      \Rule UPON arrived(X,M(profile),Y):- \\
      if(group@CS == group@profile)\\
      then do(Deliver(X,M,Y)).\label{rule-communication-receivedm}
      
  }\save

\end{footnotesize}
\end{ruleset}

In Rule~\ref{rule-communication-post}, when the user wants to publish a post to its subscribers, the controller will read local subscriber list and push the post to each of them. It will also update its attribute lastTenPosts of its profile in its control state. When the subscriber receives the post or message, according to the Rule~\ref{rule-communication-receive}, controller will show the post to the user. In the meantime, to handle the situation that the user is not online, the controller can store the post or message into local file system, or use other ways to inform users, for example by sending as email. Also, the controller can save the last several posts for the search functions. Note that if the post is of the type of management, only the managerial staff can publish it.

According to Rule~\ref{rule-communication-requestsubscribe}, any user can send a subscription request to any user. The controller will attach its profile to the request. In Rule~\ref{rule-communication-subscribereply}, when the request arrives at the user, the controller will check whether there is an access control filter in its control state. If there is not, it will add the request user to the subscriber list. If there are filters, it will examine whether this user satisfies by checking the required attributes of the profile. If the requester satisfies, the controller will add it to the subscriber list and send back the result to the request user.

If a member wants to send a direct message to a specific member, its controller will append its profile to the message, as shown in Rule~\ref{rule-communication-dm}.  When the message arrives at the controller of the destination member, it will check the group id in control state to see whether it matches the group id of the sender. If it does, the member is allowed to read the message. If the member belongs to a different group from the sender, the controller will discard the message, according to Rule~\ref{rule-communication-receivedm}.

\ss{Other Implemented Functionalities}\label{others} 

Due to lack of space, we do not cover all parts of the law for this community. However, following functionalities are very useful and important in forming a complete OSN. We discuss the general idea of them.

\p{Naming and Addressing}
When an agent joins a community, it must have a way
of naming and locating other members of the community. After all, one joins a
community only if one wishes to interact with some members.
We employ a server, called secretary, that simply acts as a naming and
locating service, negotiating with agents wishing to join the group in order for
each agent to have a unique name within that group. 
More details about this mechanism are provided by \cite{min07-1}.

\p{Search} Search capability is also necessary for an OSN.
It's relatively straightforward in the centralization, comparing to be achieved in a decentralized manner.
In decentralized OSN, one can think of two search techniques---index search and content search.
The Distributed Hash Table, which is used by most of DOSN approaches, cannot do content search.
The content search can be achieved via a gossip style search protocol---the search query initiator sends
the query to its neighbors and then the neighbors forward the query to their neighbors.
This search method is widely used by some P2P systems such as Gnutella\cite{Gnutella07url}.
Both types of search, especially the DHT, are not secure and easy to be undermined\cite{urd11},
because they need the untrusted members substantially to carry out the search correctly.
It's very vulnerable if there is no regulation imposed on each participant.
We implemented the gossip style search, and used TTL (Time To Live) and forward threshold to improve efficiency.
We can make DHT secure by implementing it by law with similar technology discussed before,
but this is beyond the scope of this paper.

\s{Related work}\label{related}

The concern about the security issues of centralized OSNs motivated several
attempts to decentralize OSNs, creating several versions of what is called
 DOSNs. PeerSoN\cite{buc09-1,buc09-2,bod11,bod12,sha11-1}, Safebook\cite{cut07-1,cut07-2,cut09},
 and LotusNet \cite{aie10-1,aie10-2} are the main attempts among others.
The basic idea underlying all these projects is that each member of the social networks keeps
the data under its own control, instead of surrendering it to a central
host. This is a necessary measure of decentralization, but it is not
sufficient. Because, as we explain in \secRef{regularity}, social network
requires some global, or communal, policy to operate under.
But none of the attempts known to us at the implementation of DOSNs provides any
means for establishing such policies.

Moreover, all these attempts adopt the substrate of DHT to implement the p2p design.
As we discussed in \secRef{others}, DHT itself is not secure under the context of heterogeneous
and distributed network and easy to be compromised. It's not able to defend some attacks
if it cannot establish certain global policy to protect it. Furthermore, DHT is incapable of performing content search.
Though some improvements or work-arounds are employed to provide limited content search,
these are way off the basic requirement of an OSN.

\s{Conclusion}\label{conclusion}
This paper addresses the risks to privacy and security posed by conventional
centralized online social networks (OSNs).
These risks, which are the consequence of centralization, seem not to be of concerns to
most of the clients of OSNs such as Facebook and
Twitter.  But they are, or should be, of serious concerns to many other current
and potential applications of OSNs.

Several recent attempts have been made to decentralize OSNs, by letting each
member of such a network keep maintaining its own data. But this DOSN approach
to decentralization is not able to establish any kind of regularity over the
social network, which is necessary for both real life social community, as well
as for OSNs.

We have introduced a decentralized architecture of OSNs, called OSC, for
\emph{online social community}, which is able to establish regularities
concerning both the membership of OSC and the manner in which its members
interact. The preliminary testing and experiments of our implementation 
show that our method is feasible and promising.

\bibliographystyle{IEEEtran} 
\bibliography{biblio,dosn}

\end{document}